# Phase-locking mechanism in non-sequential double ionization

Ivan P. Christov

*Department of Physics, Sofia University, 1164 Sofia, Bulgaria*

**Abstract:** Here, we identify a new mechanism where the early stage of electron ionization in the field of strong femtosecond laser pulse determines the final yield of non-sequential double ionization. By using simple trajectory methods we prove that the powerful short wavelength laser field causes an injection locking between its phase and the phases of the electron trajectories, which is next responsible for the enhanced or suppressed ionization for given intensities in the knee region. It is shown that both ionization and entanglement of the final electron state can be easily controlled by introducing frequency chirp of the laser field. Our methods allow one to quantify the quantum correlations due to the different mechanisms of strong field ionization.

The non-sequential double ionization is undoubtedly one of the most dramatic manifestations of electron-electron correlation taking place in the presence of powerful laser radiation. Experiments with linearly polarized sub-100 fs pulses with intensities in the range $10^{14}$-$10^{16}$ W/cm$^2$ have demonstrated ionization yields exceeding by orders of magnitude the predictions of the standard theories where the electrons are emitted independently one after the other (sequential double ionization) [1,2]. In case of two electrons (e.g. in helium) it has been clarified that during the act of non-sequential double ionization (NSDI) one of the electrons interacts strongly through its Coulomb potential with the other electron which also moves under the combined action of the nuclear potential and the external field. The re-collision mechanism where an accelerated electron returns back to the core and kicks off the other electron is held responsible for significant part of the NSDI yield which may also include re-collision excitation and subsequent field ionization [3,4]. An apparent signature of NSDI is the knee structure of the ionization yield after the laser pulse as function of the peak laser intensity, where that yield may stay almost constant, which has been observed by exact numerical simulations in reduced dimensionality models [5,6]. While the re-collision mechanism is dominant in the long wavelength high intensity regime where tunneling through the core potential occurs, for lower intensities and higher frequencies the role of the re-collision scenario is replaced by sequential and non-sequential multi-photon ionization. The increasing interest to that regime is motivated by the rapid experimental development of powerful short wavelength sources, such as the free-electron lasers, which can deliver terawatt femtosecond VUV pulses.

In this Rapid Communication we consider the NSDI in one-dimensional helium atom, where for laser wavelength 248 nm the knee structure has already been predicted to exhibit an unusual drop of the ionization yield for intensities ~6x10$^{14}$ W/cm$^2$ (see Ref. [7,8], and Fig.1(c) below). In order to further clarify the short-wavelength NSDI mechanism here we focus on the early stages of electron ionization where both electrons are still close to the core while experiencing the rising front of the laser pulse, where there is almost no tunneling. Our goal is to prove that the ionization yield after the laser pulse and hence the final NSDI output is predetermined to a large extend by the early history of the correlated electron motion. Our study is favored by the fact that at the early stage of ionization the electron motion is more tractable because it does not exhibit the near-chaotic behavior of the accelerated electrons driven by the external field.

For one dimensional helium atom with soft-core Coulomb potentials, subjected to laser field $E(t)$, the Hamiltonian in atomic units reads (e.g. in [9]):

$$H(x_1, x_2) = -\frac{1}{2}\frac{\partial^2}{\partial x_1^2} - \frac{1}{2}\frac{\partial^2}{\partial x_2^2} - \frac{2}{\sqrt{1+x_1^2}} - \frac{2}{\sqrt{1+x_2^2}} + \frac{1}{\sqrt{1+(x_1-x_2)^2}} \quad (1)$$
$$-(x_1+x_2)E(t)$$

The ground state of the atom is calculated numerically by imaginary-time propagation of the time-dependent Schrodinger equation on a two-dimensional grid, which gives ground state energy of -2.238 a.u. First we assume an electric field with six cycle trapezoidal envelope, with two cycle turn-on and turn-off. In previous work [7,8] it has been found that the double ionization signal calculated as an integral of the joined probability density for $|x_1|, |x_2| > 5$ a.u. exhibits a characteristic knee structure as function of the peak laser intensity in the range 2x10$^{14}$ W/cm$^2$ – 8x10$^{14}$ W/cm$^2$. Since for these parameters the electrons may deviate significantly from the core it is particularly useful to depict their motion by using two bundles of quantum trajectories for the two electrons, calculated from the two-body wave-function using the de Broglie-Bohm relation [10]:

$$\frac{dx_i^k(t)}{dt} = \text{Im}\left[\frac{1}{\Psi(x_i, x_j, t)}\frac{\partial \Psi(x_i, x_j, t)}{\partial x_i}\right]_{x_{i,j}=x_{i,j}^k(t)} ; i,j=1,2; k=1,...,N \quad (2)$$

where *k* enumerates the trajectories for the *i*-th electron, considered as an ensemble of *N* Monte Carlo walkers, which has been prepared to reproduce the ground state distribution given by $|\Psi(x_1,x_2,t=0)|^2$. In fact, the quantum trajectories calculated from Eq.1 and Eq.2 describe the motion of very small volumes of the time-dependent "quantum fluid" for each electron, where here we consider the NSDI trajectories only. Although these trajectories depict the electrons as distinguishable particles (e.g. as inner and outer electron) they can be very useful in evaluating the electron motion during the time evolution.

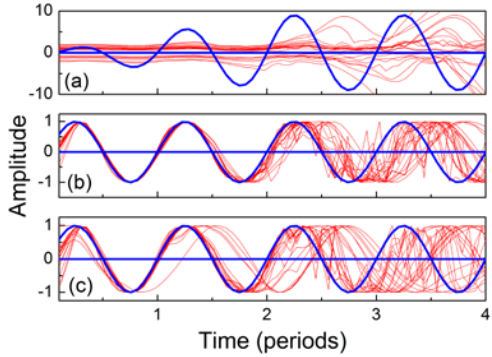

Fig. 1. Original NSDI electron trajectories -(a); trajectories after post-processing (red lines) for $4.5 \times 10^{14}$ W/cm$^2$ -(b), and for $6 \times 10^{14}$ W/cm$^2$ -(c). The blue lines depict the normalized electric field.

In order to analyze quantitatively the electron motion in a periodic external field we first calculate from Eqs.1,2 a bunch of oscillating random trajectories where during the first few cycles of the laser field these oscillations are quasi-periodic, with increasing amplitude along the leading front of the pulse (Fig1(a)). The phases of these oscillations depend not only on the phase of the laser but also on the local environment of a given trajectory, i.e. the Coulomb field. In fact, one physical reason for the phase detuning of the trajectories with respect to each other and with respect to the laser is the penetration of the electrons in the barrier formed by the coulomb potential and the external field which occurs even for higher frequencies and modifies their velocity given by Eq.(2), thus leading to transient non-adiabatic effects [11]. Therefore, spontaneous and/or driven synchronization (injection locking) of the trajectories of the different electrons during the early ionization stage may be expected, which may enhance or suppress the NSDI signal at later stages.

In order to determine the relative phases of the essentially random trajectory oscillations for the two electrons we first calculate the "envelope" of each trajectory which is next used to normalize that trajectory by dividing its waveform to its envelope. To this end we use standard procedure based on zeroing the negative-frequency part of the Fourier spectrum of a complex signal attributed to that waveform and then transform it back to time. As a result a sets of normalized complex oscillating trajectories are obtained with real parts shown in Fig.1(b),(c)), which are reminiscent of the "phase-only" oscillations used in mechanics of forces oscillators [12] where the phase is calculated as inverse tangent of the imaginary to the real part of the complex signal. Notice that since that is a part of the post-processing which we apply to the already calculated quantum trajectories, it does not in any way affect the underlying physics. Finally, we calculate the phase difference between those "phase-only" trajectories, for each separate couple which represents the two electrons. Because of the random local environment experienced by each trajectory during the first few cycles of the laser, the phases and hence the phase differences of the trajectories are also random functions of time but that randomness is exactly what determines the degree of injection-locking caused by laser field considered as an external oscillator. This is clearly seen from Fig.1(b) and (c) where for peak intensity $4.5 \times 10^{14}$ W/cm$^2$ the trajectory oscillations are less de-phased as compared to $6 \times 10^{14}$ W/cm$^2$, as they are for $8 \times 10^{14}$ W/cm$^2$ (see the green line in Fig.2(a)). Clearly, if the electrons were free they would oscillate in complete phase with the laser. In order to quantify the phase-locking of the electron trajectories we build histograms of the phase difference between the trajectories considered as random variables in time, for different peak laser intensities. In this way we are able to account for the phase locking for all times of interest, as time averages.

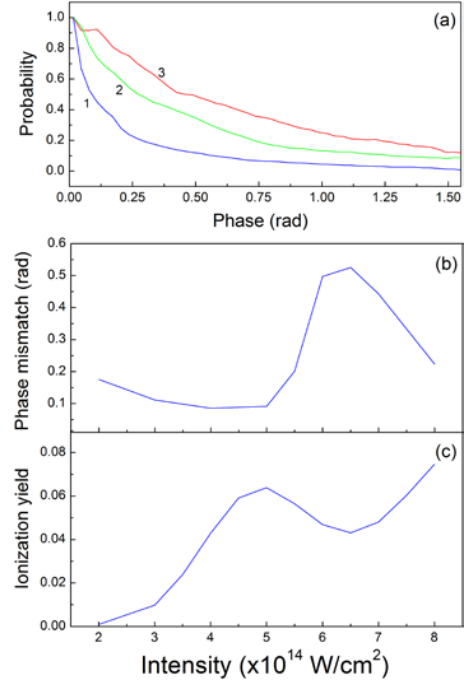

Fig. 2. (a) -Normalized smoothed histograms of the distributions of the relative phases for the two electron trajectories for different laser intensities, for the first three periods of a flat-top pulse: blue line $-4.5 \times 10^{14}$ W/cm$^2$, red line $-6 \times 10^{14}$ W/cm$^2$, green line $-8 \times 10^{14}$ W/cm$^2$; (b)-phase mismatch between the trajectories of the two electrons as function of the peak laser intensity where the knee structure of the NSDI yield in (c) is observed.

Figure 2(a) shows three selected distributions (smoothed histograms) of the relative phase of the trajectories for the two electrons, for different laser intensities, where each point along the x-axis represents a time average over the corresponding relative phases. Figure 2(b) depicts the averaged phase mismatch between the trajectories defined as the full width at half maximum of the phase distributions. It is clearly seen that the phase mismatch between the trajectories of the two electrons may vary several times between the peak and the dip of the knee in Fig.2(c), hence proving that the electron trajectories near the local ionization peak around 5x10$^{14}$ W/cm$^2$ are much better phase-locked than those near the dip around 6x10$^{14}$ W/cm$^2$, which clearly indicates that for short wavelengths the NSDI yield is closely related to the phase locking of the electron trajectories at the early ionization stages. In terms of forced oscillators the phase locked trajectories are better resonantly enhanced by the laser field considered as an external force which leads to higher ionization peak in Fig.2(c).

of the laser could directly modify the phases of the quantum trajectories and could therefore easily shift the phase-locked region (or the whole knee) to different peak intensity. Here we calculate the NSDI yield for a six cycle Gaussian pulse with and without a linear frequency chirp. The value of the chirp is limited by the spectral content of the initial transform-limited pulse such that for pulse shape given by $E(t) = E_0(t)\exp(-t^2/T^2)\cos(\omega_0 t + \gamma t^2)$ the maximum positive or negative phase modulation attainable would be $\gamma = \pm 1/(2T^2)$, while keeping the pulse energy constant. Figure 3(a) shows with black line the near-knee dependence of the NSDI yield versus laser intensity for transform-limited Gaussian pulse, with duration close to that of the flat-top pulse considered earlier, which is qualitatively close to the corresponding dependence of Fig.2(c), while the red and the green curves show that dependence for positively and negatively chirped Gaussian pulses, respectively. It is seen that, as expected, the NSDI is strongly modified by the frequency chirp of the laser where the ionization can be enhanced or suppressed for a given peak intensity. This is in support of the phase locking mechanism as seen from Fig.3(c) which shows the phase mismatch curves for the three cases where the laser chirp shifts the dip in the ionization yield by shifting the phase mismatch range.

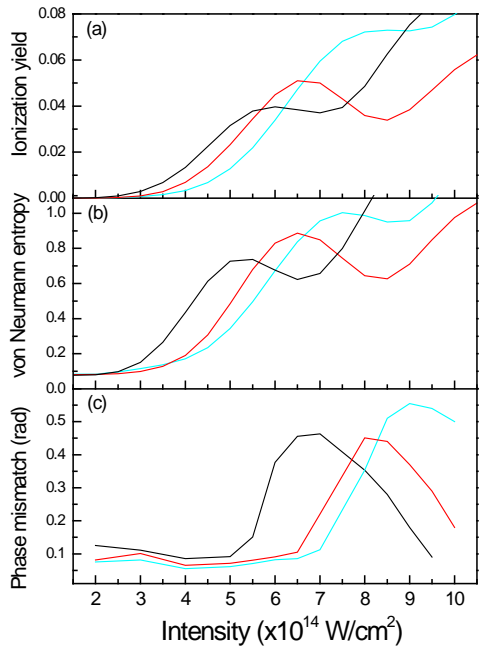

Fig. 3. NSDI yield -(a), von Neumann entropy -(b), and phase mismatch between the trajectories of the two electrons –(c) as function of the peak laser intensity, for transform limited Gaussian pulse (black lines), positive frequency modulation (red lines), and negative frequency modulation (green lines).

Having established the effect of phase locking of the electron trajectories on the ionization yield in NSDI, it is interesting to explore the possibility to control the NSDI by using phase-modulated laser field, which can be easily achieved in experiment. Such control would rely on the mechanism discussed above where the phase modulation

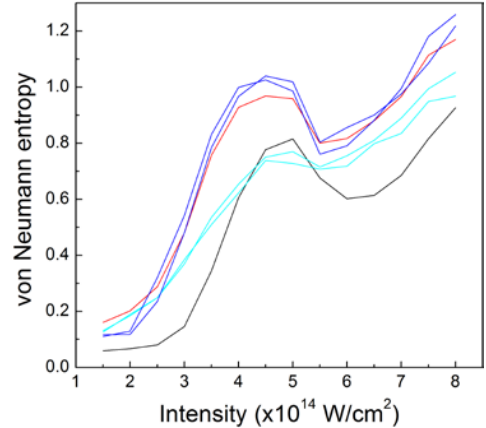

Fig. 4. Quantum Monte-Carlo predictions for the von Neumann entropy as function of the laser intensity for flat-top pulse due to: the whole trajectory ensemble (black line), NSDI-related trajectories (red line), trajectories which end up in first and third quadrants (green lines), and trajectories which end up in second and fourth quadrants (blue lines).

Since the knee region is where the ionization is most sensitive to the degree of correlation between the electrons one may use statistical measures such as the von Neumann entropy to quantify that correlation. Previously, other authors have used the inverse purity function [8] to quantify the electron correlation at the end of the laser pulse. We show in Figure 3(b) how the phase modulation of the laser pulse modifies the von Neumann entropy, which appears to be another evidence for coherent control in the strong field regime [13]. It is seen that the entropy curves qualitatively follow the ionization yield, similarly to the findings of [8].

The use of quantum trajectories offers the opportunity to explore the correlation effects for trajectories which end up in different quadrants of the 2D configuration space of the 1D helium because the different quadrants correspond to different mechanisms of electron ionization [9]. Inasmuch as the standard calculation of both von Neumann entropy $S_{vN} = -tr(\rho \ln \rho)$ [14], and inverse purity function [8], employ the one-body reduced density matrix $\rho(x, x', t) \sim \int \Psi(x, x_2, t) \Psi^*(x', x_2, t) dx_2$ calculated from the whole two-body wave function, it is difficult to distinguish the degree of correlation for specific regions of configuration space, e.g. that due to the NSDI only. To overcome that difficulty here we employ the recently proposed time-dependent quantum Monte-Carlo (TDQMC) method [15] which uses a ensembles of particles and waves as Monte Carlo walkers in physical and in Hilbert space such that these are propagated concurrently in time. That approach allows one to calculate the reduced density matrix of the interacting electron without referencing the many-body wave function [15]:

$$\rho_i(x_i, x'_i, t) = \frac{1}{N} \sum_{k=1}^{N} \varphi_i^{k*}(x_i, t) \varphi_i^k(x'_i, t) \quad ; i = 1, 2 \qquad (3)$$

where each one-body wave function $\varphi_i^k(x_i, t)$ guides a concurrent walker in physical space. The additional degree of freedom provided by the Monte-Carlo walkers allows us to restrict the summation in Eq.3 to only those wave functions for which the corresponding walkers are located in certain region along the x-axis. For example, the reduced density matrix for NSDI requires that the wave functions which contribute to Eq.3 correspond to walkers for the two electrons which simultaneously occupy regions above 5 a.u. away from the core. It should be noted that although both the density matrix of Eq.3 and the ionization yield are additive quantities, the entropy is not, and therefore we are not allowed to add the entropies due to the different ionization mechanisms. However, we may still compare the entropies calculated from Eq.3 in order to assess the electron correlation in different regimes. Figure 4 shows the von Neumann entropy predicted by the TDQMC calculation as function of the laser intensity after the flat-top pulse, where the black line represents the entropy due to the whole trajectory ensemble while the red line is the entropy due to the NSDI only. With green lines the entropies due to walkers which end up in first and third quadrant of the two dimensional configuration space are depicted, while the blue lines correspond to the walkers which end up in the second and fourth quadrants. Figure 4 clearly shows that the final entropy due to NSDI (red line) is closer to the entropy due to trajectories which end up in the second and the fourth quadrants, but it is further away from the entropy due to walkers which end up in the first and the third quadrants (green lines). That can be attributed to the Coulomb repulsion which pushes the electrons located initially in the first and the third quadrants away from each other yet at the early stages of ionization. At the same time the red and the blue curves show enhanced entanglement due to NSDI as compared to the black curve which essentially depicts the entanglement due to both sequential and non-sequential ionization. Notice that all entropies predicted by the quantum Monte Carlo calculation in Fig.4 are somewhat enhanced for higher intensities because the walkers cannot pass each other easily in 1D geometry.

In summary, by using advanced methods to analyze the propagating quantum trajectories of ionizing 1D helium atom subjected to short wavelength (248 nm) six-cycle laser pulse we were able to isolate a new mechanism which relates the final ionization yield of non-sequential double ionization to the early stages of atom-field interaction where the multi-photon ionization is dominant over the tunneling. An injection-locking mechanism was identified to be responsible for phase locking of the trajectories of the two electrons as well as between the electrons and the laser field for different regions along the ionization knee such that the two electrons may be ejected in a close sequence. That phase-locking mechanism allows one to efficiently control the ionization yield by introducing an appropriate phase modulation of the laser. An advanced quantum Monte Carlo calculation allows one to specify the von Neumann entropy as a measure of quantum correlation and entanglement in the different quadrants in the 2D configuration space. For longer wavelength (e.g. 780 nm) the re-collision mechanism of NSDI is expected to prevail, however, in that regime the electron trajectories are mostly quasi-chaotic and are therefore less tractable using the amplitude/phase analysis of this paper. As a result, for 780 nm we have found phase dependence similar to that in Fig.2(b) but it is less pronounced due to processes which may occur beyond the first few cycles of the laser. It is assumed here that there exist a range of intensities along the leading front of the laser where there is an overlap between the multi-photon and the tunneling ionization and where these two channels may compete. In fact, the drop in the knee structure for 780 nm laser pulse has been explained by classical models [16] where that drop has been attributed to the decrease of re-collision efficiency with the intensity. Finally, it was verified that the findings of this paper hold for pulses twice as long without significantly changing the phase-locking mechanism discussed.

This material is based upon work supported by the Air Force Office of Scientific Research under award number FA9550-19-1-7003.